# Optimal coupling of entangled photons into single-mode optical fibers

**R. Andrews**
*Department of Physics, Faculty of Agriculture and Natural Sciences, The University of the West Indies, St. Augustine, Republic of Trinidad and Tobago, W.I..*
*randrews@fans.uwi.tt*

**E. R. Pike and Sarben Sarkar**
*Department of Physics, King's College London, Strand, London WC2R 2LS, UK.*
*roy.pike@kcl.ac.uk, sarben.sarkar@kcl.ac.uk*

**Abstract:**

We present a consistent multimode theory that describes the coupling of single photons generated by collinear Type-I parametric down-conversion into single-mode optical fibers. We have calculated an analytic expression for the fiber diameter which maximizes the pair photon count rate. For a given focal length and wavelength, a lower limit of the fiber diameter for satisfactory coupling is obtained.

©2004 Optical Society of America

**OCIS codes:** (190.0190) Nonlinear optics; (190.4410) Parametric processes

### References and links

1. D. C. Burnham and D. L. Weinberg, "Observation of simultaneity in parametric production of optical photon pairs," Phys. Rev. Lett. **25**, 84-87 (1970).
2. Z. Y. Ou, X. Y. Zou, L. J. Wang and L. Mandel, "Experiment on nonclassical fourth-order interference," Phys. Rev. A **42**, 2957-2965 (1990).
3. C. K. Hong and L. Mandel, ''Theory of parametric frequency down-conversion of light,'' Phys. Rev. A **31**, 2409-2418 (1985).
4. Z. Y. Ou and L. Mandel, "Violation of Bell's inequality and classical probability in a two-photon correlation experiment," Phys. Rev. Lett. **61**, 50-53 (1988).
5. P. G. Kwiat, K. Mattle, H. Weinfurter, A. Zeilinger, A. V. Sergienko, and Y. H. Shih, Phys. Rev. Lett. **75**, 4337-4341 (1995).
6. C. H. Bennett, G. Brassard, C. Crepeau, R. Jozsa, A. Peres, and W. K. Wootters, "Teleporting an unknown quantum state via classical and Einstein-Podolsky-Rosen channels," Phys. Rev. Lett. **70**, 1895-1899 (1993).
7. S. L. Braunstein and H. J. Kimble, "Teleportation of continuous quantum variables," Phys. Rev. Lett. **80**, 869-872 (1998).
8. L. Vaidman, "Teleportation of quantum states," Phys. Rev. A **49**, 1473-1476 (1994).
9. D. Bouwmeester, J-W Pan, K. Mattle, M. Eibl, H. Weinfurter and A. Zeilinger, "Experimental quantum teleportation," Nature **390**, 575-579 (1997).
10. J-W Pan, D. Bouwmeester, H. Weinfurter and A. Zeilinger, "Experimental entanglement swapping : Entangling photons that never interacted," Phys. Rev. Lett. **80**, 3891-3894 (1998).
11. D. Bouwmeester, J-W Pan, M. Daniell, H. Weinfurter and A. Zeilinger, Phys. Rev. Lett. **82**, 1345-1349 (1999).
12. A. K. Ekert, J. G. Rarity, P. R. Tapster, and G. M. Palma, "Practical quantum cryptography based on two-photon interferometry," Phys. Rev. Lett. **69** (9), 1293-1295 (1992).
13. W. Tittel, J. Brendel, H. Zbinden, and N. Gisin, "Quantum cryptography using entangled photons in energy-time Bell states," Phys. Rev. Lett. **84** (20), 4737-4740, 2000.
14. C. Kurtsiefer, M. Oberparleiter, and H. Weinfurter, "High-efficiency entangled photon pair collection in type-II parametric fluorescence," Phys. Rev. A **64**, 023802 (2001).
15. S. Castelletto, I. P. Degiovanni, M. Ware, and A. Migdall, "Coupling efficiencies in single photon on-demand sources," ArXiv:quant-ph/0311099 (2003).
16. F. A. Bovino, P. Varisco, A. M. Colla, G. Castagnoli, G. D. Giuseppe, and A. V. Sergienko, "Effective fiber coupling of entangled photons for quantum communication," Optics Communications **227**, 343-348 (2003).
17. E. R. Pike and S. Sarkar, *The Quantum Theory of Radiation,* Oxford University Press (1995).
18. R. Andrews, E. R. Pike and Sarben Sarkar, "Photon correlations and interference in type-I optical parametric down-conversion," J. Opt. B: Quantum semiclass. Opt. **1**, 588-597 (1999).

19. A. Yariv, *Optical Electronics in Modern Communications* (5[th] Edition, Oxford Series in Electrical and Computer Engineering).
20. J. W. Goodman, *Introduction to Fourier Optics* (McGraw-Hill, New York, 1968), chaps. 7 and 8.
21. Q. Cao and S. Chi, "Approximate Analytical Description for Fundamental-Mode Fields of Graded-Index Fibers: Beyond the Gaussian Approximation," Journal of Lightwave Technology, Volume 19, Issue 1, 54- (2001).
22. A. Mair, A. Vaziri, G. Weihs and A. Zeilinger, "Entanglement of the orbital angular momentum states of photons," Nature, vol. 412, 313-316 (2001).
23. A. L. Migdall, D. Branning, and S. Castelleto, "Tailoring single-photon and multiphoton probabilities of a single-photon on-demand source," Phys. Rev. A, **66**, 053805-1-053805-4 (2002).

## 1. Introduction

The optical process of spontaneous parametric down-conversion (SPDC) involves the virtual absorption and spontaneous splitting of an incident (pump) photon in a transparent nonlinear crystal producing two lower-frequency (signal and idler) photons [1-3]. The pairs of photons can be entangled in a multi-parameter space of frequency, momentum and polarization. In type-I SPDC the photons are frequency-entangled and the signal and idler photons have parallel polarizations orthogonal to the pump polarization. Entangled photons have been used to demonstrate quantum nonlocality [4,5], quantum teleportation [6-8] and, more recently, quantum information processing [9-11] and quantum cryptography [12,13]. One of the challenges in recent practical schemes such as quantum cryptography and quantum communication is to maximize the efficiency of coupling of single photons into single-mode optical fibers. Several models and experiments have been developed to predict and measure the coupling efficiencies of down-converted light into single-mode fibers [14,15,16]. In the model and experiment of *Kurtsiefer et. al.* [14], the angular distribution of non-collinear type-II down-converted light for a given spectral bandwidth is calculated. Their idea is to match the angular distribution of the photon pairs to the angular width of the fiber mode. *Bovino et. al.* [16] produced a more rigorous model in which the dependence of the coupling efficiency on crystal length and walk-off were investigated. The importance of hyper-entanglement in type-I down-conversion and its possible use as a resource in the field quantum information has been recently discussed [22]. Single photon on-demand sources have also been designed using an array of type-I down-converters [23]. In contrast to previous work, we present a detailed theoretical model describing single-photon mode coupling in the simple situation of collinear type-I down-conversion. The pair photon count rate is calculated and an analytic expression which determines the condition for optimum single-photon coupling in single-mode optical fibers is obtained in terms of experimental parameters.

## 2. Amplitude for pair detection in single-mode fibers

The amplitude for detecting photon pairs at conjugate space-time points $(\vec{r}_1, t_1)$ and $(\vec{r}_2, t_2)$ is defined by

$$A^{(2)} = \left\langle \vec{E}_H^+(\vec{r}_1, t_1) \vec{E}_H^+(\vec{r}_2, t_2) \right\rangle \qquad (1)$$

where $t_1$ and $t_2$ are detection times of signal and idler photons and $\vec{E}_H(\vec{r}_i, t)$ are the Heisenberg electric field operators [18]. In the steady state the right-hand-side of Eq. (1) can be expressed as

$$\int d^3r_3 \int d^3k_1 \int d^3k_2 U^*_{\vec{k}_1\lambda_1}(\vec{r}_3) U^*_{\vec{k}_2\lambda_2}(\vec{r}_3) U_{\vec{k}_0\lambda_0}(\vec{r}_3) f_p(\vec{r}_3) \left(\frac{\hbar\omega_{k_0}}{2\varepsilon_0}\right)^{\frac{1}{2}} \left(\frac{\hbar\omega_{k_1}}{2\varepsilon_0}\right)^{\frac{1}{2}} \left(\frac{\hbar\omega_{k_2}}{2\varepsilon_0}\right)^{\frac{1}{2}}$$

$$\times \langle \alpha_{k_0}, 0 | \vec{E}_I^{(+)}(\vec{r}_2, t_2) \vec{E}_I^{(+)}(\vec{r}_1, t_1) | \alpha_{k_0}, k_1, k_2 \rangle \delta(\omega_{k_1} - \omega_{k_2} - \omega_{k_0})$$

(2)

$\vec{E}_I(\vec{r}_i, t)$ are the interaction-picture electric field operators at some arbitrary but specified detection points $\vec{r}_1$ and $\vec{r}_2$ and $f_p(\vec{r}_3)$ is a function which describes the shape of the pump in the transverse direction; $U_{\vec{k}_i\lambda_i}(\vec{r}_3)$ are plane-wave modes describing the electromagnetic field in the crystal with $\vec{k}_1, \vec{k}_2$ as the wave vectors of the signal and idler photons, $\vec{k}_0$ is the wave vector of the incident pump photon and $\lambda_i$ are polarization indices. (As in our previous studies we have not introduced the effects of a change in the linear refractive index between the crystal and its surroundings. The incorporation of this difference does not qualitatively change our conclusions). We have therefore assumed that the nonlinear crystal is embedded in a medium whose linear refractive is the same as that of the crystal. Optical fibers used in the analysis are assumed to have the same refractive index as the crystal.

The initial state of the electromagnetic field $|0, \alpha_{k_0}\rangle$, consists of a coherent state with wave-vector $k_0$ (the monochromatic pump beam) with other modes in the vacuum state $|0\rangle$. We take the quantized electric field in the fiber as

$$\vec{E}^+(\vec{r}, t) = i \sum_{\lambda\lambda'} \int d^3k' \int d^3k \left(\frac{\hbar\omega_k}{2\varepsilon_0}\right)^{1/2} \vec{\varepsilon}_{\vec{k}\lambda} \beta_{\vec{k}'\lambda', \vec{k}\lambda} U^f_{\vec{k}\lambda}(\vec{r}) e^{-i\omega t} a_{\vec{k}\lambda} \quad (3)$$

where the coupling coefficient $\beta_{\vec{k}'\lambda', \vec{k}\lambda}$ is a projection onto the appropriate fiber mode and is defined as [19]

$$\beta_{\vec{k}'\lambda', \vec{k}\lambda} = \int dx dy U^{in}_{\vec{k}'\lambda'}(x,y) U^{f*}_{\vec{k}\lambda}(x,y) \quad (4)$$

$U^{in}_{\vec{k}'\lambda'}(\vec{r})$ are spatial modes incident on the fiber and $U^f_{\vec{k}\lambda}(\vec{r})$ are the fiber modes. For single-mode fibers we assume a spatial mode profile that is independent of $\vec{k}$ and $\lambda$. We may therefore suppress the fiber mode indices in Eq. (4) to simplify the notation. After performing the integration over the volume of the crystal and evaluating the expectation value, Eq. (2) simplifies to

$$\int d^3k_1 \int d^3k_2 \tilde{f}_p(\vec{k}_{1t} + \vec{k}_{2t}) \text{sinc}\left(\Delta_{k_{1z}k_{2z}} \frac{d}{2}\right) \beta_{k_1\lambda_1} \beta_{k_2\lambda_2} U^f(\vec{r}_1) U^f(\vec{r}_2)$$

$$\times \left(\frac{\hbar\omega_{k_1}}{2\varepsilon_0}\right)^{1/2} \left(\frac{\hbar\omega_{k_2}}{2\varepsilon_0}\right)^{1/2} e^{-i\omega_{k_1}t_1} e^{-i\omega_{k_2}t_2} \delta(\omega_{k_0} - \omega_{k_1} - \omega_{k_2})$$

(5)

where $d$ is the length of the crystal, $\Delta_{k_{1z}k_{2z}} = k_0 - k_{1z} - k_{2z}$ and the signal/idler wave vectors are defined as $\vec{k}_i = \hat{x}k_{ix} + \hat{y}k_{iy} + \hat{z}k_{iz}$ where $i = 1,2$ denote signal, idler modes and $k_{ix}, k_{iy}, k_{iz}$ are components of the wave vector along the x,y,z directions respectively. $\tilde{f}_p(\vec{k})$ is the two-dimensional Fourier transform of $f_p(\vec{r})$. For degenerate signal and idler photons emitted at a cone angle $\theta^*$ to the pump and for frequencies close to the degenerate frequency, the sinc function in Eq. (5) can be approximated to unity in a first-order analysis. If we assume that the divergence of the pump is negligible over the length of the crystal, i.e., the crystal is sufficiently thin, and that the transverse shape of the pump is gaussian, we can take the shape function as

$$f(\vec{r}_t) \propto \exp\left(-\frac{x^2 + y^2}{w_p}\right) \quad (6)$$

Thus, $\tilde{f}_p(\vec{k}) \propto \exp(-\frac{1}{4} w_p^2 k^2)$, where $w_p$ is the effective width of the pump beam. In terms of photon frequencies, a first-order analysis gives

$$\tilde{f}_p(\vec{k}_{1t} + \vec{k}_{2t}) \propto \exp(-\frac{1}{4} w_p^2 v^2 (\omega_{k_1} - \omega_{k_2})^2 \sin^2\theta^*) \quad (7)$$

where $v$ is the first-order dispersion coefficient of the nonlinear crystal. The collinear situation is obtained for $\theta^* = 0$.

## 3. Calculation of the coupling coefficient

The coupling coefficient is defined in Eq.. (4) as the overlap integral of the input modes and the fiber mode. For our calculations in the collinear geometry the fibers are positioned along the z-axis. In experiments one needs to use a beam splitter to separate the beams as is shown in Fig. 1. A lens of focal length $f$ is used to focus collinear signal and idler beams onto identical single-mode fibers through a 50/50 beam splitter. The pair count rate is then measured by single-photon detectors D1 and D2 connected to a correlator.

Since the beam splitter introduces constant phase factors into the pair detection amplitude, our calculations will also be valid for this experiment.

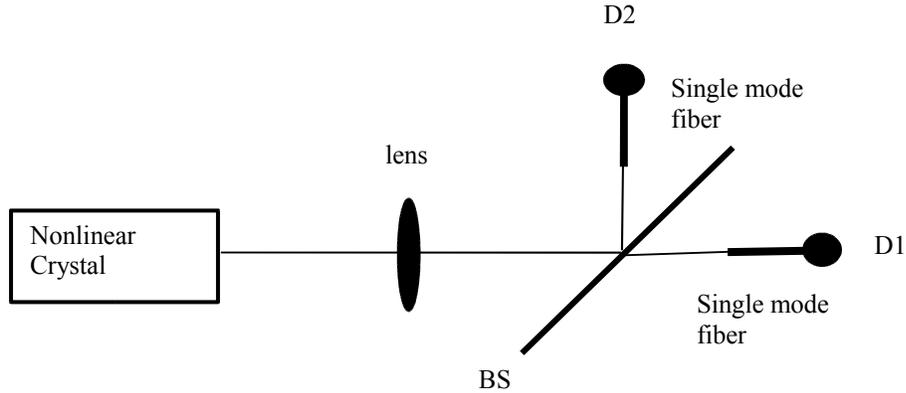

Fig. 1: Schematic of the experimental setup to determine pair photon coupling in fibers

Assuming that the input face of the fibers is at the image plane of the lens and that plane-wave modes, described by the free-field interaction-picture electric field operators in Eq. (2), are incident on the lens, we can employ diffraction theory to obtain the electric field modes at the image plane of the lens [20]. We obtain

$$U_{k\lambda}^{in}(x,y) = \frac{1}{ikd'} e^{ik(d'+d)} e^{i\frac{k}{2d'}(x^2+y^2)} \int_0^{2\pi}\int_0^R r d\theta dr \left( e^{i\left(\frac{k}{2d'}-\frac{k}{2f}\right)r^2} e^{-ik\left(\frac{x\cos\theta}{d'}+\frac{y\sin\theta}{d'}\right)r} \right) \tag{8}$$

where $d$ is the distance from the center of the crystal to the lens and $d'$ is the distance from the lens to the image plane. We are also assuming that a circular region of the lens of radius $R$ is being irradiated by the down-converted light. We take the fiber modes, $U^f(x,y)$, as Gaussian modes [21] defined by

$$U^f(x,y) = \frac{1}{w_0\sqrt{\pi}} \exp\left(-\frac{x^2+y^2}{2w_0^2}\right) \tag{9}$$

where $w_0$ is the radius of the fiber mode. Substituting Eq. (8) and Eq. (9) in Eq. (4) we obtain the coupling coefficient $\beta_{k\lambda}$ as

$$\beta_{k\lambda} \propto \frac{w_0^2 d'^2 + ikw_0^4 d'}{kd' w_0(d'^2+k^2 w_0^4)} \frac{\left(1-e^{(-A_k+iB_k)R^2}\right)}{(-A_k+iB_k)} \tag{10}$$

where $A_k$ and $B_k$ are defined as

$$A_k = \frac{k^2 w_0^2}{2(d'^2+k^2 w_0^4)}$$

$$B_k = \left(\frac{1}{2d'}-\frac{1}{2f}\right)k - \frac{k^3 w_0^4}{2d'(d'^2+k^2 w_0^4)} \tag{11}$$

**4. Two-photon count rate in single-mode fibers**

Consider the situation in which the fiber is in the focal plane of the lens, i.e., $d' = f$. We assume that Gaussian filters are used to select frequencies close to the degenerate frequency. Then for $k^2 w_0^4 \ll f^2$ and to first-order in the dispersion of the crystal, the two-photon amplitude in Eq. (5) approximates to

$$A^{(2)} \propto \int_{-\infty}^{\infty} e^{-i\tau x - \frac{x^2}{\Delta^2}} \left(1-e^{(a+bx)}\right)^2 dx \tag{12}$$

where

$$a = \left(-\frac{k^{*2} w_0^2}{2f^2} - i\frac{k^{*3} w_0^4}{2f^3}\right) R^2$$

$$b = \left(-\frac{k^* w_0^2 n_g}{cf^2} - i\frac{3}{2}\frac{n_g k^{*2} w_0^4}{2cf^3}\right) R^2 \tag{13}$$

$\tau$ is the time difference between the detection of signal and idler photons, $k^*$ is the degenerate phase-matched wave number, $n_g$ is the group refractive index and $\Delta$ is the bandwidth of the down-converted light reaching the fibers. To obtain the count rate we take the modulus square of the integral in Eq. (12). Typical photon detectors have resolving times much larger than the coherence time of the down-converted photons. The limits in the integration with respect to $\tau$ may then be taken as effectively from $\tau = -\infty$ to $\tau = \infty$.
The $\tau$ integration gives a delta function which reduces a double integral to a single $x$-integral. After performing the $x$-integral and using the approximation $k^2 w_0^4 << f^2$ for a typical bandwidth $\Delta \approx 10^{12}$ rads$^{-1}$, we obtain, to a good approximation,

$$C \propto \left(1 - 4e^{-y} + 6e^{-2y} - 4e^{-3y} + e^{-4y}\right) \qquad (14)$$

where $C$ is the pair count rate and $y = \dfrac{k^{*2} w_0^2 R^2}{2f^2}$. This approaches a maximum when $y \geq 4.4$. This gives the fiber radius $w_0 \geq 1.5 \dfrac{f\lambda}{\pi R}$ for optimal coupling where $\lambda$ is the wavelength in the fiber. For a fiber diameter equal to the Rayleigh width of the diffraction pattern of the aperture, i.e., $w_0 = \dfrac{f\lambda}{\pi R}$, the two-photon count rate in the fiber is only 56% of the maximum.

## 5. Conclusion

We have presented, from a first-principles calculation, a fully quantized multimode theory to describe the coupling of single photons generated by collinear Type-I SPDC into single-mode optical fibers. We have thus produced a practical recipe to maximize the coupling of single photons into optical fibers. As pointed out earlier this is important for quantum cryptography and quantum communication applications. For the type-I collinear down-conversion studied here, we find no significant dependence of the coupling on the transverse width of the pump beam and crystal length which have been obtained for non-collinear geometries. The important parameters for the collinear case are simply the photon wavelength, the focal length of the lens and the fiber diameter.